\documentclass{icrc2009}

\usepackage{graphicx}   
\usepackage[caption=false]{caption}    
\usepackage[font=footnotesize]{subfig} 
\usepackage{fixltx2e}
\usepackage{url}

\newcommand{\shorttitle}[1]%
{\markboth{Proceedings of the 31\MakeLowercase{$^{st}$} ICRC, {\L}\'{o}d\'{z} 2009}{#1} }
\newcommand{\etal}{\MakeLowercase{\textit{et al. }}} 

\begin{document}
\title{Gamma-ray and Cosmic Ray Astrophysics from 10 TeV to 1 EeV with
       the large-area ($>$\,10\,km$^2$) air-shower Detector SCORE}

\author{\IEEEauthorblockN{M. Tluczykont\IEEEauthorrefmark{1}, T. Kneiske\IEEEauthorrefmark{1}, D. Hampf\IEEEauthorrefmark{1}, D. Horns\IEEEauthorrefmark{1},\\
\IEEEauthorblockA{\IEEEauthorrefmark{1}Department Physik, Universit{\"a}t Hamburg, Luruper Chaussee 149, 22761 Hamburg}}}

\author{\IEEEauthorblockN{M. Tluczykont, T. Kneiske, D. Hampf, D. Horns,\\
\IEEEauthorblockA{Department Physik, Universit{\"a}t Hamburg, Luruper Chaussee 149, 22761 Hamburg}}}
\shorttitle{Tluczykont \etal UHE Astronomy with SCORE}
\maketitle

\begin{abstract}
We propose to explore the so-far poorly measured cosmic ray and gamma-ray sky (accelerator sky) in the energy range from 
10\,TeV to 1\,EeV.
New physics questions might be addressed 
in this last remaining observation window of gamma-ray astronomy. The very high beam-energies provided by Cosmic 
accelerators and the air-shower detection technique naturally imply an entanglement between fundamental questions 
of astroparticle physics and particle physics.
The new large-area (10\,km$^2$) wide-angle (1\,sr) air Cherenkov detector SCORE (Study for a Cosmic ORigin Explorer)
is based on non-imaging Cherenkov light-front sampling with sensitive 
large-area detector modules of the order of 1\,m$^2$. The lateral photon density and arrival-time 
distribution will be sampled up to large distances from the shower core.
The physics motivations, the detector concept and first simulation results will be presented.
\end{abstract}

 
\section{The last Gamma-ray Observation Window}
\begin{figure}[!t]
  \centering
  \includegraphics[width=3.0in]{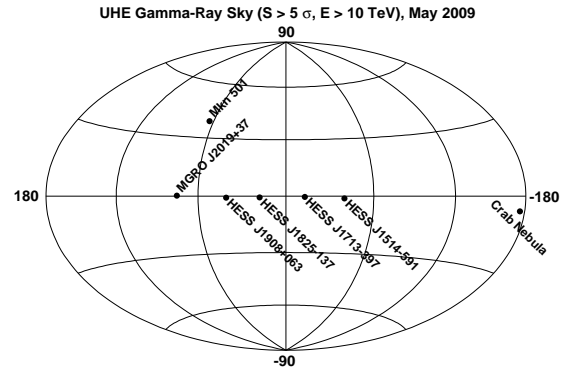}
  \caption{The Ultra High-Energy Gamma-Ray Sky in Galactic coordinates.
           Shown are sources so far detected with more than 5\,$\sigma$
           above E$>$10\,TeV. Status as of May 2009.}
  \label{tluczykont_score_uhegammasky}
 \end{figure}

%
%
%
The sensitivity level of existing and currently planned gamma-ray detectors
is optimized for the very high-energy regime (VHE, 100\,GeV\,$<$\,E\,$<$\,30\,TeV).
The sensitivity to the ultra-high energy gamma-ray regime (UHE, E\,$>$\,30\,TeV)
is limited because previously, the trend in development of detectors for
gamma-ray astronomy was dominated by the focus on low energy
thresholds.
So far, the number of sources detected above 10\,TeV is less than 10 (see Figure~\ref{tluczykont_score_uhegammasky}),
and above 100\,TeV no detection has been reported so far.
A number of fundamental physics questions need to be addressed in the UHE gamma-ray regime.
Future and planned experiments such as CTA \cite{hofmann:cta}, HAWC \cite{sinnis:2005a} or LHAASO \cite{cao:lhaaso} will improve
the situation in the UHE gamma-ray regime.
However, due to dropping event statistics with rising energy,
the key to UHE gamma-ray astronomy is a very large instrumented
area of the order of 10 to 100\,km$^2$.
While such large instrumented areas seem impracticable using the well-established
imaging air Cherenkov technique (order of 10000 channels/km$^2$),
the non-imaging air Cherenkov technique provides a complementary possibility
that comes along with some advantages, such as a small number of channels per km$^2$
(few 100 per km$^2$) and a wide field of view (order of sr).
%
We have started the development of SCORE (Study for a Cosmic ORigin Explorer),
a 10\,km$^2$ prototype detector
for a ground-based wide-angle large-area air-shower detector
for non-imaging gamma-ray astronomy and cosmic ray physics from 10\,TeV to 1\,EeV.
With a wide field of view of the order of 1\,sr (continuously monitoring
a large part of the sky) and
a focus on the highest energies, this project is complementary to
(yet independent of) existing and planned experiments.
SCORE is understood as a first step towards a 100\,km$^2$ detector of similar 
design, H$_i$SCORE (Hundred Square-km Cosmic Origin Explorer).

In addition to its capabilities for gamma-ray observations, a large area air-shower
array such as SCORE is the ideal detector for cosmic ray physics, allowing
observations of primary cosmic rays in the domain of
transition between the Galactic and the extra-galactic component.
The detection technique of SCORE is based on
shower front sampling and timing with Cherenkov light (non-imaging technique).
{SCORE provides a unique approach to measure the entire
longitudinal development using the shower-front arrival-time distribution
(at distances from the shower core $>$100\,m), allowing
detailed spectral and composition measurements,
and allowing gamma-hadron separation via reconstruction of the
shower-depth.}
With an instrumented area of 10\,km$^2$, SCORE covers an energy range
for measurements of the primary cosmic ray spectrum and the chemical
composition from 100\,TeV -- overlapping with direct
cosmic ray observations -- to
1\,EeV\footnote{The number of detected cosmic rays above 1\,EeV is expected to be 10 per year} -- overlapping with fluorescence measurements.

In the next section the main physics objectives of SCORE will be addressed,
followed by a description of the detector principle (section~\ref{tluczykont_score_score})
and first simulation results (section~\ref{tluczykont_score_results}).
The event reconstruction is described in a dedicated contribution to this conference
\cite{hampf:2009a}.

\section{Physics Objectives}
\label{tluczykont_score_physics}
\subsection{Cosmic Ray Origin}
The energy range from 10 TeV to several 100 TeV is the most natural 
energy regime for a solution of the almost 100-year-old mystery of the origin of cosmic rays.
Gamma-rays are emitted at the site of cosmic ray acceleration via hadronic interactions
of the cosmic rays with ambient matter, producing neutral pions subsequently decaying 
into gamma-rays (and neutrinos from charged pions).
At least some Galactic accelerators -- the pevatrons -- 
are expected to accelerate cosmic rays via shockfront acceleration \cite{fermi:1949a}
up to knee energies ($\approx$\,4\,PeV).
The gamma-ray spectral energy distribution of a pevatron is expected
to reach up to several 100\,TeV\footnote{The cut-off energy in gamma-ray spectra is
roughly a factor of 10 lower than in the underlying hadronic
spectrum.}
(see Figure~\ref{tluczykont_score_sensitivity}).
An additional spectral hardening (flattening) below the cut-off energy is predicted
in diffusive shock acceleration theory when non-linear
effects are taken into account \cite{voelk:2002b,berezhko:1006}.

In the VHE regime, an ambiguity in gamma-ray spectra arises from the fact that gamma-rays can also be
produced by leptonic acceleration processes (inverse Compton scattering).
Therefore, if observations only provide spectra up to few tens of TeV, the underlying production
mechanism of a gamma-ray signal needs to be
resolved with precise spectral measurements over a wide range in energy (multi-wavelength)
and detailed comparisons to predictions from hadronic
and leptonic scenarios.
Luckily, the ambiguity of gamma-ray spectra disappears
in the UHE regime, where the inverse Compton scattering cross-section
enters the Klein-Nishina regime, loosing efficiency.
This results in a softening (i.e. steepening) of gamma-ray spectra from leptonic accelerators
beyond 10\,TeV. As opposed to that, a hard (i.e. flat)
gamma-ray spectrum continuing up to few hundred TeV
would be a clear signature of hadronic acceleration.
For comparison, the observed spectral energy distribution of 
the shell-type supernova remnant RXJ\,1713.7-3946 \cite{aharonian:rxj1713}
is shown as a dotted line in Figure~\ref{tluczykont_score_sensitivity}.
The spectrum of this supernova remnant -- as well as of any other object so far detected in 
the VHE regime -- can still be accomodated within leptonic acceleration scenarios. 
Comparing the mesured spectra with expected pevatron spectra shows that
an expectedly existing population of PeV accelerators might have been 
missed out completely by VHE experiments, even more so when considering that 
such accelerators might be widely extended structures larger than the field of
view of the most sensitive instruments (IACTs, 3--5$^\circ$).
Recent detections of partly unidentified sources (H.E.S.S., Milagro)
yield hard gamma-ray spectra that extend up to few tens of TeV.
These results underline the necessity to carry out observations in the UHE regime.
As opposed to previous experiments, SCORE is specifically designed for the UHE gamma-ray regime and
for spatially extended emission.
With the SCORE experiment, we aim at identifying cosmic ray accelerators by searching for
hard spectra reaching up to few 100\,TeV, therwith contributing to the resolution of
the mystery of the origin of cosmic rays.
The observation of UHE gamma-ray emission from extended gas clouds illuminated by 
nearby cosmic ray sources, might open up a possibility to map the Galactic
cosmic ray acceleration efficiency.

\subsection{Propagation and Absorption}
At such high energies as considered here,
gamma-rays are attenuated via e$^+$e$^-$ pair production with the photons of the Galactic
interstellar radiation fields (IRF) and with the Cosmic Microwave Background (CMB).
It was shown that for Galactic objects the attenuation
reaches a maximum around 100\,TeV from the Galactic IRF and at 2\,PeV from the CMB
\cite{moskalenko:2006a}.
While the former depends on the Galactic longitude and local radiation fields, the latter is universal.
If the distance of the observed gamma-ray sources is known, the density of the IRF might
be inferred from the strength of the absorption by pair production (or from spectral features).
Inversely, Galactic absorption might also open up a new possibility
to infer distances from the measurement of gamma-ray spectra, if
the IRF in the line of sight is known. Such a new method for distance estimation of Galactic
objects might also be possible if the universal absorption by the CMB could be measured.
The expected attenuation by pair production might be altered by photon/axion
conversion (e.g. \cite{steffen:axions}). Photons produced
at the source travelling through the Galactic magnetic field might convert into axions propagating
without absorption. If a reconversion of these axions back into photons happens before arriving at earth,
the photon signal would appear to be stronger than expected.
The same effect could arise if photon/hidden-photon oscillations \cite{zechlin:2008a} would occur.
Another effect that might alter the expected absorption by pair production is
the modification of the e$^+$e$^-$ pair production threshold in case of
Lorentz invariance violation.

\subsection{The Local Supercluster}
Ultra-high-energy cosmic rays from the local super-cluster are expected to interact with the 
CMB, initiating intergalactic pair cascades that could be observable in gamma-rays.
The emission is expected to match the large scale of the local super-cluster structure
and would be measurable as anisotropic diffuse emission in the total field of view of the experiment.
Alternativeley, the accelerators of ultra-high-energy cosmic rays might exhibit point-like
emission or halo-like structures resulting from the interactions of the accelerated particles
with the surroundings of the source.


\subsection{Cosmic Ray Composition}
By measuring the longitudinal development of hadronic air-showers (using the shower-front arrival-time distribution),
detailed spectral and cosmic ray composition measurements are possible from 100 TeV to 1 EeV.
SCORE will detect $2\times10^7$ cosmic rays per year above 1\,PeV and 10 cosmic rays per year above 1\,EeV.
Furthermore, the reconstruction of the 
shower depth will allow an independent measurement of the inelastic proton-proton cross-section overlapping and exceeding 
LHC energies.

 
\section{The Detector Concept}
\label{tluczykont_score_score}
The detector principle is based on the shower-front sampling technique
using Cherenkov light. A large array of wide-angle light-sensitive detector
stations is used. The concept of the detector modules used as working
assumption for the SCORE detector was adapted from previous gamma-ray
experiments, such as
THEMISTOCLE \cite{fontaine:1990a,behr:themis},
HEGRA AIROBICC \cite{karle:1995a}, or BLANCA \cite{cassidy:blanca}.
Similar detector modules are also used in the currently being built
TUNKA array for cosmic-ray physics \cite{budnev:tunka}.
As compared to the previous experiments, three aspects of SCORE will
be different: an instrumented area larger by more than an order of magnitude,
larger detector station areas and larger inter-station spacing
(see Table~\ref{tluczykont_score_properties}).
\begin{table*}[th]
\caption{\label{tluczykont_score_properties}Basic design characteristics of the SCORE detector
and for the possible extension H$_i$SCORE. 
Corresponding values for TUNKA, AIROBICC and BLANCA are also given.}
\centering
\begin{tabular}{|l|l|l|l|l|l|l|}\hline
                        & \bf SCORE             & H$_i$SCORE    & TUNKA         & BLANCA        & AIROBICC & THEMISTOCLE    \\\hline
Instrumented area $A [km^2]$      	& \bf 10                & 100           & 1             & 0.2   	& 0.04     &    0.08\\
Detector station area $a [m^2]$    		& \bf 0.5--1.5          & 0.5--1.5      & 0.2           & 0.1   	& 0.13     &    0.5\\
Field of View $FoV [sterad]$      	& \bf 0.84              & 0.84          & 1.8           & 0.12  	& 1        &    0.008\\
Inter-station distance $d [m]$   		& \bf 100--200          & 100--200      & 85            & 35    	& 30       &    50--100\\
Number of detector stations 			& \bf O(100)            & 2601          & 133           & 144   	& 49       &    18\\\hline
\end{tabular}
\end{table*}
Advances made in technology allow improvements
to the original detector components such as
improved photo-sensitive detectors,
fast trigger and readout electronics, therewith allowing
a measurement of the Cherenkov photon arrival time distribution.

\begin{figure}[!t]
  \centering
  \includegraphics[width=3.0in]{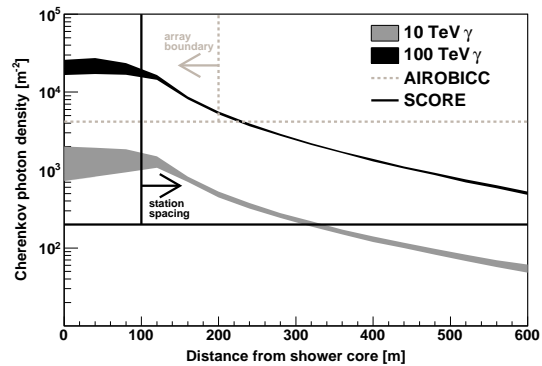}
  \caption{Lateral photon density function (LDF) of Cherenkov light on sea level for airshowers
           initiated by a 10\,TeV gamma-ray shower (grey area) and a 100\,TeV gamma-ray shower.
           The sensitivity level of one SCORE detector station is indicated by the solid line.
           For comparison, the corresponding sensitivity level for AIROBICC is also shown (dashed grey box).}
  \label{tluczykont_score_lateral}
 \end{figure}

A very large instrumented area (10\,km$^2$) is required 
to reach sufficiently large event statistics at ultra-high energies,
and is achieved with a low array density, i.e. large inter-station spacings.
A reasonable value for the detector station spacing can be
derived from the lateral photon density function (LDF) of Cherenkov light at
observation level, shown in Figure~\ref{tluczykont_score_lateral}.
Within a radius of 120\,m around the shower core position
the LDF is roughly constant, but shows
a large spread from shower to shower. Fluctuations are much lower
beyond 120\,m. 
With the envisaged station spacing of 100--200\,m, SCORE will primarily 
be sensitive to the outer part of the LDF (promising good reconstruction quality),
as illustrated with the solid lines in Figure~\ref{tluczykont_score_lateral}.
The low photon density far away from the shower core justifies
the chosen large individual detector station areas.
For comparison, the corresponding sensitive range of the AIROBICC experiment is also shown
(dashed lines), illustrating the novel approach chosen with SCORE.
Another important aspect for the reconstruction with the SCORE detector 
is the usage of the full timing information
from the arrival time distribution of the Cherenkov photons at each detector
station.
The event reconstruction of SCORE is based on the combination of information from
the lateral photon density distribution and the arrival time distribution of Cherenkov
photons (see \cite{hampf:2009a}, this conference). 

Each detector station consists of four photomultiplier tubes (PMTs) equipped with four
light-collecting Winston cones of 30$^\circ$ half-opening angle pointing to the zenith.
All four modules (PMT+cone), including the trigger, readout electronics and communication
are planned to be encased in a box equipped with a sliding lid.
The advantages of using four PMT channels per station are the possibility
to suppress false triggers from nightsky background (NSB) light by a local coincidence trigger
condition and the requirement of a large light collecting area ($a$).
A total area of $a$~=~0.5\,m$^2$ can be achieved when using four 8'' PMTs and a Winston cone height
of 0.5\,m.
%
The final design of the Winston cones is work in progress. Different solutions are envisaged for the
cone structure (plastic mold, segmented metal sheets) and for the reflective inner surface
(vaporization, reflective foils).
A fast signal readout and digitization in the GHz regime are needed. Different solutions 
such as analog ring samplers or domino ring samplers (DRS) are under study.
We are currently testing usage of the DRS\,4 chip that was developed by the PSI\footnote{\url{http://midas.psi.ch/drs}}
and which is also used by the MAGIC experiment.

The envisaged trigger scheme encompasses two levels.
A local station trigger is issued when the sum of all four PMT signals passes a given
threshold (few $\sigma$ above NSB level).
In the second level, a two-fold station coincidence is required from a sub-array of
detector stations (3$\times$3 or 4$\times$4) and the signals are then sent to the
data acquisition. Individual sub-arrays overlap to avoid dead trigger zones.
Alternatively a smart station concept could be implemented, issuing a trigger
based on a next-neighbour station condition.
For a large array such as SCORE, with large inter-station spacings, relative timing is
a critical issue.
A relative timing accuracy of better than 5\,ns is required to limit
the impact on reconstruction (mainly angular resolution).
For time synchronization we plan to use optical fibers for clock distribution
to the local station clocks.

An extension of the concept we studied so far is to equip the underside of each
sliding lid of the station boxes with scintillator material.
With such a setup, continuation of data taking during daytime would be
possible using SCORE as a very large charged particle shower front sampling array.

\section{First Simulation Results}
\begin{figure}[!t]
  \centering
  \centering\includegraphics[width=3.0in]{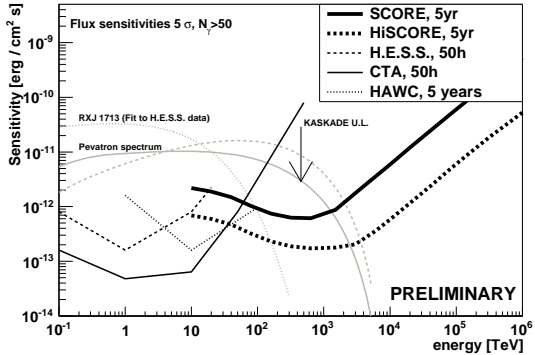}
  \caption{Sensitivity of SCORE (solid line) after 5 years of operation.
        Further sensitivities are given for H.E.S.S., CTA and HAWC.
        For comparison, the expected pevatron spectrum (exponential cut-off energy was 4\,PeV,
        linear acceleration: solid grey line,
        non-linear acceleration: dashed grey line)
        and a fit to H.E.S.S. data from the shell-type SNR RXJ\,1713.7-3946 (dotted grey line),
        as well as an upper limit by the KASCADE experiment \cite{antoni:2004a} are given.
          }
  \label{tluczykont_score_sensitivity}
 \end{figure}

Air showers were simulated using CORSIKAv675 \cite{heck:1998a}.
The detector simulation (\emph{sim\_score}) was implemented
on the basis the \emph{iact} package \cite{bernloehr:2008a:iact}.
We included signal pulse shaping and used basic assumptions on PMT quantum
efficiency and Winston cone reflectivity.
In order to estimate the sensitivity of the SCORE detector, the effective trigger area 
was calculated for mono-energetic gamma-rays and protons. These results were used to estimate
the number of signal events (for a pevatron-like spectrum as in Figure~\ref{tluczykont_score_sensitivity})
and background events based on the parametrization of the cosmic ray spectrum by \cite{hoerandel:2003a}.
An energy-dependent angular resolution as found in simulations (see \cite{hampf:2009a}, this conference)
and a Quality factor for gamma-hadron separation of 1.5 were assumed.
\label{tluczykont_score_results}
The point-source sensitivity of SCORE was calculated for a detection of at least 50 gamma-rays 
at a significance level of 5\,$\sigma$ in 5 years (see Figure~\ref{tluczykont_score_sensitivity}, thick solid line).
The SCORE sensitivity is competitive with existing and currently planned instruments
around 100\,TeV and significantly extends the energy range in the UHE gamma-ray regime.

\section{Conclusion and Outlook}
With the developement of the SCORE detector we aim at opening up the last remaining
observation window of gamma-ray astronomy, the UHE gamma-ray regime. Furthermore,
cosmic ray spectral and chemical composition measurements from 100\,TeV to
1\,EeV will be possible with SCORE. First simulation results show a competitive sensitivity
above 100\,TeV and demonstrate a good potential for new physics in the UHE regime.

With further improvements of reconstruction and gamma/hadron separation power, we
hope to improve the expected sensitivity of the SCORE detector.
With a larger array of 100\,km$^2$ (H$i$SCORE, thick dashed line Figure~\ref{tluczykont_score_sensitivity}), the sensitivity could further be enhanced.

\end{document}